\documentstyle[psfig]{article}
\begin{document}
\begin{center}
{\Large\bf Experimental Characterization of the Ising Model
in Disordered Antiferromagnets}

\vspace{0.25in}

{\large D. P. Belanger}\\
{\small Department of Physics, University of California,
Santa Cruz, CA 95064 USA}\\
\end{center}
\begin{abstract}

The current status of experiments on the $d=2$ and $d=3$
random-exchange and random-field Ising models, as
realized in dilute anisotropic antiferromagnets, is discussed.
Two areas of current investigation are emphasized.  For
$d=3$, the large random field limit is being investigated
and equilibrium critical behavior is being characterized
at high magnetic concentrations.

\end{abstract}

\section{Introduction}

The Ising model is one of the most studied and basic
models for phase transitions.
In this article, the current status of experimental studies
characterizing two classic models of second-order
phase transitions in short-range interaction systems
in the presence of quenched
disorder, the random-exchange Ising model (REIM)
and the random-field Ising model (RFIM), is
presented.  The discussion concentrates on experiments
in dilute, insulating, anisotropic antiferromagnets,
the systems that have yielded the best understood data
for these two models.  The REIM is realized in zero magnetic field and
the RFIM with a field applied along the spin-ordering direction.
The REIM is rather well characterized
experimentally, theoretically and through computer
simulations.  The $d=2$ RFIM is fairly well
characterized, although the scaling behavior of scattering
near the destroyed phase transition is still being investigated.
The understanding of the RFIM for $d=3$ is not as complete,
though significant progress has been made in
the past few years, and it is this model that will be the
main focus of this short review.  The early history of the $d=3$ RFIM was
fraught with controversial interpretations of the data,
a result of severe nonequilibrium effects.  Nevertheless,
some experimental groups realized from the start that underlying
the observed, complicated behavior is a new kind of phase
transition.  Efforts to characterize the new critical behavior
were thwarted by the severe nonequilibrium effects.  These nonequilibrium
effects have recently been overcome by going to sufficiently
high magnetic concentration and a complete
characterization of the universal $d=3$ RFIM critical behavior is
possible and underway.  The most recent static critical behavior
will be compared to results from computer simulations and theory.
In addition to these low-field behaviors, much has been learned about the
high-field limit of the RFIM.  An overview will be given of
the phase diagram and the different behaviors observed.

Experiments
have been performed on REIM and RFIM systems for more than two
decades.  Since the experiments performed some
time ago have been reviewed previously\cite{by91}, 
they will be included here only as needed to give a perspective
on the current physical understanding of the models.
Theory and simulation results will be included as needed
for the interpretation of the experiments.
Another classic model of ordering in the presence of
disorder, the spin glass, will be covered in another
review\cite{nordblad00} in this Ising Colloquium and so will not
be discussed here in detail, although some spin-glass-like
behaviors do occur at low magnetic concentrations and at
high magnetic fields.

Table 1 shows the most frequently measured static critical behaviors
associated with a phase transition.  We will make reference to
the universal parameters defined in Table 1 as needed.

\begin{table}[t]
\begin{center}
\begin{tabular}{||l|c||}
\hline\hline
specific heat&$C=A^{\pm}|t|^{-\alpha}+B$\\
\hspace{0.3in}for $\alpha \rightarrow 0$&$C=A\ln |t|$\\
order parameter ($T<T_c$)&$M_s=M_o|t|^{\beta}$\\
fluctuation correlation length&$\xi=\xi_o ^{\pm}|t|^{-\nu} = 1/\kappa$\\
staggered susceptibility&$\chi_s=\chi_o^{\pm}|t|^{-\gamma}$\\
disconnected susceptibility&$\chi_s^d=\chi_o^{d\pm}|t|^{-\bar{\gamma}}$\\
\hline\hline
\end{tabular}
\end{center}
\caption[Asymptotic critical behavior]
{The asymptotic forms for commonly measured static critical
behaviors.  The superscript + (-) on the amplitudes signifies $T>T_c$
($T<T_c$).  The exponent values and the amplitude ratios
are universal quantities that depend only on the general properties
of the system.}
\label{table:neut-seven}
\end{table} 

\section{Experiments on Pure $d=2$ and $d=3$ Anisotropic Antiferromagnets}

Observations of asymptotic static critical behavior in the pure
$d=2$ and $d=3$ Ising antiferromagnets are very well
documented.  The magnetic specific heat ($C_m$) critical behavior has been
characterized using the optical linear birefringence
techniques\cite{fg84,bkj84} on the $d=2$
$\rm {Rb}_2\rm {CoF}_4$ system\cite{fkjcg83}
and the $d=3$ $\rm {FeF}_2$ system\cite{nbkji83}.
The measured pure Ising critical exponents
$\alpha$ and the amplitude ratios $A^+/A^-$
are in superb agreement with very many theoretical and simulation
results.  The birefringence technique is
particularly useful and more accurate than pulsed specific heat
techniques since it is insensitive to the large
phonon contributions which are particularly difficult to
handle for $d=2$.

The critical behavior of the staggered susceptibility and
correlation length have been determined with
neutron scattering for
$\rm {K}_2\rm {CoF}_4$\cite{chb84} for $d=2$ and in
$\rm {FeF}_2$\cite{by87} for $d=3$.
In general, the scattering line shapes for the pure and REIM
systems away from the Bragg scattering point follow the scaling
behavior of the spin-spin correlation function
\begin{equation}
{\chi _s^{}}(q)=A^{\pm}\kappa ^{\eta -2}f(q/\kappa) \quad ,
\end{equation}
where $\gamma = \nu (2-\eta)$.
For both $d=2$ and $d=3$, the scaling functions used in data
analysis are approximate ones\cite{tf75,fb67} that differ
significantly from the mean-field (MF) Lorentzian
\begin{equation}
f(q/\kappa)= \frac {1}{1+(q/\kappa)^2} \quad ,
\end{equation}
as can be seen in Fig.\ 1 where various scaling functions are compared.
The deviation from the Lorentzian is more pronounced for $d=2$
and for $T<T_c(H)$ in both dimensions. 

The order parameter critical behavior has been determined
using neutron scattering\cite{chb84} in the $d=2$ compounds.   The
M\"{o}ssbauer technique\cite{wb67} was used in the study of the $d=3$ system.

The results for the pure Ising model in $d=2$ and $d=3$ are
summarized in Table 2.  Note that the Rushbrooke scaling relation
\begin{equation}
2\beta + \gamma + \alpha \ge 2
\end{equation}
is satisfied as an equality for both cases.  Included in Table 2
are the results from a few theoretical and simulation studies.
No attempt is made to review the vast literature on the pure Ising models.

\begin{table}[]
\begin{center}
\begin{tabular}{||l|c|c||}
\hline\hline
Pure $d=2$&Experiment& Theory\cite{o44}\\
\hline
$\alpha$    & $0.00 \pm 0.01$\cite{nbkji83} & $O(\log |t|)$\\
$A^+/A^-$&$1.01\pm 0.00$\cite{nbkji83}&$1(\log |t|)$\\
$\beta$ & $0.155 \pm 0.02$\cite{chb84} & $1/8$\\
$\nu$ & $1.02 \pm 0.05^+$\cite{chb84} & $1$\\
&$1.12\pm 1.13^-$\cite{chb84}&\\
$\kappa ^+/\kappa -$&$0.54 \pm 0.06$\cite{chb84}&$1/2$\\
$\gamma $ & $1.82 \pm 0.07$\cite{chb84} & $7/4$\\
&$1.92 \pm 0.20$\cite{chb84}&\\
$\chi^+/\chi^-$&$32.6 \pm 3.7$\cite{chb84}&$37.33$\\
\hline
Pure $d=3$&Experiment& Theory\\
\hline
$\alpha$    & $0.11 \pm 0.005$\cite{bnkjlb83} & $0.1099\pm 0.0007$\cite{cprv99}\\
&&$0.109 \pm 0.004$\cite{gz98}\\
$A^+/A^-$&$0.54 \pm 0.02$\cite{bnkjlb83}&$0.55$\cite{bgz74}\\
$\beta$ & $0.325 \pm 0.005$\cite{wb67} & $0.32648 \pm 0.00018$\cite{cprv99}\\
&&$0.3258 \pm 0.0014$\cite{gz98}\\
$\nu$ & $0.64 \pm 0.01$\cite{by87} & $0.63002 \pm 0.00023$\cite{cprv99}\\
&&$0.6304 \pm 0.0013$\cite{gz98}\\
$\kappa ^+/\kappa -$&$0.53\pm 0.01$\cite{by87}&$0.52$\cite{bgz74}\\
$\gamma $ & $1.25 \pm 0.02$\cite{by87} & $1.2371 \pm 0.0004$\cite{cprv99}\\
&&$1.2396 \pm 0.0013$\cite{gz98}\\
$\chi^+/\chi^-$&$4.6\pm 0.2$\cite{by87}&$4.8$\cite{bgz74}\\
\hline\hline
\end{tabular}
\end{center}
\caption[Critical exponents for the $d=3$ pure Ising model]
{
The pure $d=2$ and $d=3$ Ising static critical exponents obtained from
experiments, theory and Monte Carlo
simulations.
}
\end{table}

\section{Random-Exchange Experiments in Dilute $d=2$ and $d=3$ Anisotropic Antiferromagnets}

The REIM is realized in dilute, anisotropic insulating antiferromagnets
when the site dilution does not result in strongly frustrated
bonds (that would lead to spin-glass behavior).
Random-exchange phase transitions are observed in $d=2$ and
$d=3$ systems and these appear to be in good accord with theory
and simulations.

The $d=2$ REIM $C_m$ critical behavior was observed
using the birefringence technique\cite{fkjcg83} on the magnetically dilute
antiferromagnet $\rm {Rb}_2\rm {Co}_{0.85}\rm {Mg}_{0.15}\rm {F}_4$.
The approximately logarithmic divergence is compatible with
theoretical predictions\cite{s00,fhy00}.
The scattering critical behavior\cite{hcni87} of the
compound $\rm {Rb}_2\rm {Co}_x\rm {Mg}_{1-x}\rm {F}_4$ was analyzed
using approximate scattering line shapes\cite{tf75,fb67} developed for
the pure $d=2$ Ising model.  The successful analysis using
these line shapes suggests that the correct line shape is
close to the pure one.  The static critical behavior
of the $d=2$ Ising model is quite well characterized
by experiments and theory as shown in Table 3.

\begin{table}[]
\begin{center}
\begin{tabular}{||l|c|c||}
\hline\hline
$d=2$ Random & Experiment & Theory\cite{dd83,s00} \\
Exchange & ($H=0$)  &  \\ 
\hline
$\alpha$    & $\approx O(\log |t|)$\cite{fkjcg83}& $O(\log (\log 1/|t|))$\\
$A^+/A^-$&$0.95\pm 0.10$\cite{nbkji83}&$1(\log |t|)$\\
$\beta$ & $0.13 \pm 0.02$\cite{hcni87}&$1/8$\\
$\nu$&$1.08 \pm 0.06^+$\cite{hcni87}&$1$\\
&$1.58\pm 0.52^-$\cite{hcni87}&\\
$\kappa ^+/\kappa -$&$0.98 \pm 0.02$\cite{hcni87}&$1/2$\\
$\gamma $ & $1.75 \pm 0.07$\cite{hcni87} & $7/4$\\
&$2.6 \pm 0.6$\cite{hcni87}&\\
$\chi^+/\chi^-$&$19.1\pm 5.0$\cite{hcni87}&37.33\\
\hline
$d=3$ Random & Experiment & Theory \\
Exchange & ($H=0$)  &  \\
\hline
$\alpha$    & $-0.10 \pm 0.02$\cite{sb98}& $-0.051 \pm 0.013$\cite{bfmspr98}\\
$A^+/A^-$&$1.55\pm 0.15$\cite{wb00}&$-0.5$\cite{n83}\\
$\beta$ & $0.350 \pm 0.009$\cite{rklhe88}& $0.3546 \pm
0.0028$\cite{bfmspr98}\\
$\nu$ & $0.69 \pm 0.01$\cite{bkj86}& $0.6837 \pm 0.0053$\cite{bfmspr98}\\
$\kappa ^+/\kappa -$&$0.54 \pm 0.06$\cite{bkj86}&$0.83$\cite{n83}\\ 
$\gamma $ & $1.31 \pm 0.03$\cite{bkj86}  & $1.342 \pm 0.010$\cite{bfmspr98}\\
$\chi^+/\chi^-$&$2.8 \pm 0.2$\cite{bkj86} &$1.7$\cite{n83} \\
\hline\hline
\end{tabular}
\end{center}
\caption[Critical exponents for the $d=3$ REIM]
{
The $d=2$ and $d=3$ REIM Ising static critical exponents obtained
from experiments, simulations and theory.
}

\end{table}

The $d=3$ REIM is similarly well characterized with
birefringence, neutron scattering and M\"{o}ssbauer experiments employing
$\rm {Fe}_x\rm {Zn}_{1-x}\rm {F}_2$ with the results shown in Table 3
along with some theoretical and simulation results with which they agree
very well.  The critical behavior of the specific heat of the $d=3$ REIM,
measured with birefringence techniques\cite{sb98}, is shown in Fig.\ 2.
Monte Carlo simulations\cite{bbICM}
based on the $\rm {Fe}_x\rm {Zn}_{1-x}\rm {F}_2$
system are shown in Fig.\ 3.
The birefringence technique yields a negative specific heat exponent
$\alpha$ as predicted\cite{h74}, consistent with a universality
class different from the pure Ising model where $\alpha$ is positive.
Note that, just as in the pure case, the birefringence technique
is consistent with pulsed specific heat techniques,
though the latter technique suffers from greater concentration
gradient sensitivity\cite{bkfj87} and the large phonon specific heat component.
The critical behaviors of the staggered susceptibility and
correlation length were determined from neutron scattering
experiments\cite{bkj86}.  The order parameter critical behavior
was determined from M\"{o}ssbauer studies\cite{rklhe88}.
The scattering line shape scaling
functions are not known from theory and were therefore determined
directly from the scattering data
in neutron scattering experiments\cite{sbf99} using
$\rm {Fe}_{0.93}\rm {Zn}_{0.07}\rm {F}_2$.
The results shown in Fig.\ 1 clearly indicate that the scaling
functions are fairly close to those of the pure $d=3$ case.

The REIM universal static critical parameters for $d=2$ and $d=3$ are
shown in Table 3 along with theoretical and simulation results.  In both
dimensions the agreement is excellent.  Note that the Rushbrooke
scaling relation (Eq.\ 3) is satisfied as an equality for the REIM.

\begin{table}[]
\begin{center}
\begin{tabular}{||l|c|c||}
\hline\hline
$d=3$ Random & $Fe_xZn_{1-x}F_2 $ & Monte Carlo \\
Field & ($H>0$) &  \& Exact Ground State\\
\hline
$\alpha$    & $0.0 \pm 0.02$\cite{sb98}& $ -0.5 \pm 0.2$\cite{r95}\\
&&$-0.55 \pm 0.2$\cite{nue97}\\
$\beta $ & not measured\cite{rkjr88} & $0.00 \pm 0.05$\cite{r95}\\
&&$0.02 \pm 0.01$\cite{nue97}\\
&&$0.25 \pm 0.03$\cite{bbICM}\\
$\nu$ & $0.88 \pm 0.05$ & $1.1 \pm 0.2$\cite{r95}\\
&&$1.14 \pm 0.10$\cite{nue97}\\
$\gamma $ & $1.58 \pm 0.13$ & $1.7 \pm 0.2$\cite{r95}\\
&&$1.5 \pm 0.2$\cite{nue97}\\
$\bar{\gamma}$ & $2\gamma = 3.16 \pm 0.26$ & $3.3 \pm 0.6$\cite{r95}\\
&&$3.4 \pm 0.4$\cite{nue97}\\
\hline\hline
\end{tabular}
\end{center}
\caption[Critical exponents for the $d=3$ RFIM]
{
The $d=3$ RFIM Ising static critical exponents obtained from experiments,
simulations and theory.
}

\end{table}

\section{The $d=3$ Magnetic Percolation Threshold Concentration}

As the magnetic percolation threshold concentration, $x_p$, is approached
from above in zero field, the equilibrium phase transition is expected
to approach zero temperature. For
a particular magnetic structure, $x_p$ depends on what interactions exist
between the different neighboring spins.  For example,
for $\rm {Fe}_x\rm {Zn}_{1-x}\rm {F}_2$,
$x_p=0.245$ provided only the dominant interaction between
the body-center and corner spins is considered\cite{se64}.
Except very close to $x_p$, the much smaller interactions
can be ignored.  Close to $x_p$, however,
the smaller interactions may drastically affect the behavior
and even prevent ordering above $x_p$ when they
frustrate the dominate interaction.
Both the extremely slow dynamics near percolation\cite{h85} and the
sensitivity to tiny frustrating interactions\cite{syp79} can cause the system
to exhibit spin-glass-like behavior.

A good deal of effort has focused on the properties of the system
$\rm {Fe}_x\rm {Zn}_{1-x}\rm {F}_2$ for $x$ near $x_p$.
This system has a small frustrating interaction\cite{bb00}.
The spin-glass-like properties were first elucidated
in experiments by
Montenegro, et al.\cite{rmcbp88,mcr89,mjmmr92,mlcr90,acvammrc91,mrc88}.
For $H>0$ and $x=x_p$, there exists a boundary
that resembles a de Almeida-Thouless boundary with curvature
$T-T_o \sim H^{2/\phi}$ where $\phi=3.4$, a typical spin-glass value.
It was shown with neutron scattering\cite{by93}
that there is no antiferromagnetic long-range ordering below
this boundary.  Much of the behavior
is very reminiscent of a canonical spin glass.
The detailed behavior of this sample has been studied
experimentally\cite{rmcbp88,jdnb97} and extensively modeled
in local mean-field\cite{rcm95,rc98} and Monte Carlo simulations\cite{brc00}.

\section{$d=2$ Random-Field Behavior}

Scaling arguments for Zeeman and domain wall energies
by Imry and Ma\cite{im75} as well as
considerations by Binder\cite{b83} leave little doubt that
the $d=2$ Ising transition is destroyed by the introduction of arbitrarily
small random fields.  Birefringence\cite{fkjcg83} and
neutron Bragg scattering\cite{bkj85a} experiments bear this out;
no sharp phase transition is observed in equilibrium,
though the rounded transition exhibits the expected scaling behavior.
The equilibrium region is separated from a lower temperature region
of strong hysteresis observed in the difference between data obtained upon
heating after cooling in zero field to low temperatures and then applying
the field (ZFC) and upon simply cooling the sample in the field (FC).
The boundary separating these regions is time-scale
dependent\cite{kjmsd85}.  
The domain dynamics induced with the application of a magnetic field
as well as those remaining after the field is removed at low temperatures
have been studied experimentally\cite{kjbb98,bkj85}
and theoretically\cite{snu98}.

\section{$d=3$ RFIM behavior for $x_p<x<x_e$ at low fields.}

The behavior for concentrations between $x_p$
and the percolation threshold concentration for vacancies,
$x_e =1-x_p$, with a relatively small applied
field occupied the bulk of early experimental efforts\cite{by91}.
Much of the controversy over interpretations of experimental
data involved this region of concentration and fields.  
As a result of the equivalence\cite{c84,fa79} of the dilute anisotropic
antiferromagnet in small fields and the random-field
ferromagnet often studied theoretically, it was believed
that concentrations near $x=0.5$ would yield strong
random-field effects in reasonably small fields and would
be the best realizations of the $d=3$ RFIM for phase transition
studies.  At the time
of the first experiments\cite{r81,bkj82,bkjc83,ycsbgi82},
it was generally believed, based on many theoretical arguments,
that no phase transition would be observed.  Indeed, early
neutron scattering FC experiments, by Yoshizawa, et al.\cite{ycsbgi82},
seemed to bear this out.  In particular,
a resolution-limited Gaussian Bragg peak does not occur
upon FC, although subsequent experiments\cite{bkj85}
show that the samples retain long-range order below
the phase boundary if ZFC. In contrast, the
first $C_m$ studies\cite{bkjc83,lk84}
yielded compelling evidence for a fundamentally new phase
transition governing the behavior.  The phase boundary
$T-T_c(H)\sim H^{2/\phi}$ behaves as predicted\cite{a86} with
$\phi=1.42\pm 0.03$ for random-exchange to random-field crossover\cite{fkj91}.

Despite the sharp $C_m$ peak, many have argued against it as
evidence of the existence of a phase transition.
The most recent of these is
the ``trompe l'oeil transition'' phenomenological model\cite{bfhhrt95}.
Among the assumptions of this model are that the birefringence
and $C_m$ experiments do not yield the same
behavior, the uniform magnetization is reflected by
the square of the staggered magnetization, and conventional
scaling is inoperative.
This phenomenological model was shown to be
inconsistent\cite{bkm96} when all available data are considered.
In contrast, more proven and conventional techniques of analyzing
the experimental data, as described in this review, have been
very fruitful in providing consistent results in a
meaningful scaling context.

It is clear
that the phase transition underlying the behavior is obscured
by nonequilibrium behavior below the boundary, $T_{eq}(H)$, lying
just above the phase transition and scaling in the same manner,
albeit with a slightly larger amplitude\cite{kjbr85}.  The equilibrium behavior
above $T_{eq}(H)$ can be used to extrapolate
the scattering data to the obscured phase transition boundary.
When done carefully, the boundary determined in this way
coincides with that determined via $C_m$ experiments,
which are much less sensitive to the non-equilibrium behavior that
distorts the neutron scattering data.  For a long time, the
nonequilibrium experiments represented the best random-field
results available.  One of the particularly interesting
predictions\cite{r95} of the change
in the critical behavior induced by the random fields is
that the order parameter critical exponent $\beta$ should
decrease from $0.35$ to a value near zero.  This can only be measured
below $T_c(H)$, i.e.\ in the nonequilibrium region, so it was not
clear what would be observed.  Experiments on thin films
were made\cite{bwshnlrl96} for $x=0.52$, well below
$x_e \approx 0.76$.  Not surprisingly, the results were
peculiar.  The curvature of the Bragg intensity versus $T$
was such that it would require $\beta >> 0.5$, which is hard
to justify theoretically.  Magnetic x-ray scattering data showed
similar behavior near surfaces in bulk samples, though they
were interpreted under the
``trompe l'oeil'' phenomenology and the Bragg scattering was not
separated from the fluctuation scattering\cite{bfhhrt95}.
For some years
it appeared that the problems of metastability below $T_{eq}(H)$ could
not be avoided, i.e.\ that they were intrinsic to the random-field behavior
as realized in dilute antiferromagnets.  However, insight
into the origins of the metastable domains finally led
to experiments\cite{sbf99} at high magnetic concentration as a way to
avoid the nonequilibrium behavior, as discussed later.

The metastable domains formed upon FC are themselves
quite interesting and their dynamics were studied in some
detail both experimentally\cite{lsbho92,lkf88,pkb88,lk84,mdn00}
and through simulations\cite{nu91,hb92,hbkn92}.

\section{RFIM behavior  for $x_p<x<x_e$ at high fields.}

The general phase diagram features, shown in Fig.\ 4 for the
RFIM at high fields were investigated in pioneering pulsed-field
magnetization measurements\cite{kjsmd83}
in $\rm {Fe}_{x}\rm {Zn}_{1-x}\rm {F}_2$.  Low temperature
single spin flips and the phase boundary are shown in Fig.\ 5 and,
interestingly, the behavior of the upper phase boundary appears to be
different for $x<x_e$ and $x>x_e$.  This is consistent with the
differentiation of the behaviors observed in neutron scattering experiments
above and below $x_e$.

In recent years, it has become clear that weak RFIM (small
applied field) and strong RFIM regimes exist for
$x_p<x<x_e$.
The $\rm {Fe}_{0.31}\rm {Zn}_{0.69}\rm {F}_2$ system exhibits\cite{mkjhb91}
typical low-field behavior for $H<1.5$~T, with $T_N-T_c(H)$
and $T_N-T_{eq}(H)$ scaling as $H^{2/\phi}$ with $\phi \approx 1.4$.
At larger fields, however, the curvature for $T_N-T_{eq}(H)$
changes to $\phi \approx 3.4$, a value close to that observed
in spin glasses.  Some of these features were suggested
qualitatively\cite{sgrl85,ab87}.
While the lower region
has been shown to have antiferromagnetic long-range order upon
ZFC\cite{bmmkj91}, no long-range antiferromagnetic
order is observed at the higher fields below $T_{eq}(H)$.  Instead,
spin-glass-like behavior is observed.  This is clearly the same
type of behavior observed for all fields at the percolation
threshold concentration\cite{jnm?}.  The same type of distinctive low and high
field behaviors have been observed for $x$ as large as $0.60$.
The two regions are separated by an equilibrium boundary\cite{mltl98,mltl99},
as observed for $x=0.56$ (Fig.\ 6) and $0.60$, and it
appears to decrease towards
the $H=0$ boundary at finite temperature well below $T_c(H)$ and approach
the phase transition line at finite field, a point separating the sharp
transition observed at low field and the more glassy transition at higher
fields.  This also is consistent\cite{rfm00} with specific heat peaks that are
very sharp at low fields and quite rounded at high fields\cite{sktk88,bfhhrt95}.

The distinction between high- and low-field behavior behavior
is observed as well in ac susceptibility experiments.
At low fields,
there exists a single peak which seems to be associated with
extremely slow dynamics, either from
activated dynamics or at least power-law behavior with a very
large dynamic exponent\cite{kmj86,nkj91}.
There is little hysteresis between the ZFC and FC procedures
at low $H$.
At larger fields the peak splits in the ZFC procedure only, with
a sharp peak at slightly lower temperatures than the
broader peak\cite{bkk95,rfm00}.  This splitting appears
to be associated with the
upper region of the phase diagram corresponding to the spin-glass-like
behavior\cite{rfm00}.  
These effects have not been investigated for $x>x_p$.  The high field
region for these concentrations
is still an open area for research.

\section{$d=3$ RFIM Equilibrium Critical Behavior for $x>x_e$}

It was, of course, realized very early that the metastable
domain walls at low magnetic concentrations took advantage
of vacancies.  What was not fully appreciated 
was that the Imry-Ma domain wall energy
argument\cite{im75} is not applicable when domain walls can
to a great extent pass through vacancies, avoiding the
energy cost of breaking magnetic bonds.
With sufficient vacancies, i.e.\ for $x<x_e$, domain walls
can take advantage of vacancies
to such an extent that the domain
wall energy can be insignificant.
Interestingly, every experiment that could
have detected low temperature hysteresis, particularly
neutron scattering and capacitance\cite{kjbr85} experiments, was done
for $x \le 0.72$, which is below $x_e=0.76$.
Higher concentrations were avoided since
the generated random fields are small, resulting in
relatively very narrow asymptotic random-field critical regions
around $T_c(H)$.  Nevertheless, concentrations well above $x_e$
are necessary to study the equilibrium critical behavior
and require high magnetic fields and very fine temperature
resolution.

Figure 2 shows the specific heat data for
$\rm {Fe}_{0.93}\rm {Zn}_{0.07}\rm {F}_2$,
measured with optical linear birefringence.  The specific
heat was also measured to demonstrate that, in agreement with
theory\cite{fg84} and contrary to the so-called `trompe l'oeil'
phenomenology\cite{bfhhrt95}, the data from the both techniques yield the
same critical parameters.  An important advantage of the birefringence
technique is its relative insensitivity to the concentration
gradients that tend to smear the transition\cite{bkfj87}.  
Interestingly, the critical behavior appears to be very similar
to that of lower concentrations where metastable domains
dominate the scattering behavior.  The specific heat was studied
using Monte Carlo simulations\cite{bbICM} based on the
$\rm {Fe}_{0.93}\rm {Zn}_{0.07}\rm {F}_2$ system and the results
are shown in Fig.\ 3.  Although the simulations are not of sufficient
resolution to extract the critical exponent, the similarities in
the shapes of both the REIM and RFIM indicate the same qualitative
change from the asymmetric cusp at $H=0$ to the nearly symmetric
peaks at $H>0$.  This result, however, is not seen in all MC
simulations\cite{r95}.  

The neutron scattering experiments on
$\rm {Fe}_{0.93}\rm {Zn}_{0.07}\rm {F}_2$ 
show no hysteresis below the phase transition, in stark
contrast with samples with $x<x_e$.  There is no evidence
that domains form upon ZFC or FC in this concentration
range and the line shapes are independent of the thermal
cycling procedure, implying equilibrium conditions.

Neutron scattering experiments on this sample were difficult
to analyze since the RFIM line shape is not known from theory.
In general, two different scaling functions are involved with
the form
\begin{equation}
{\chi _s^{}}(q)=A^{\pm}\kappa ^{\eta -2}f(q/\kappa) + B^{\pm}\kappa ^{\bar\eta -4}g(q/\kappa)\quad .
\end{equation}
with two independent sets of critical behavior exponents.
However, simulations\cite{bfmspr98} and
high temperature series expansions\cite{gaahs96} strongly suggest a simpler
scenario.  The predictions are that the exponents are
simply related, $\bar \eta$ being twice $\eta$, and the new
scaling function $g(q/\kappa)$ is, to a good approximation,
the square of $f(q/\kappa)$.

Not only were the universal RFIM critical parameters obtained in this
experimental study, but the scaling analysis yielded the 
spin-spin correlation scaling function $f(q/\kappa)$.
This scaling function is compared to several other known spin-spin
correlation scaling functions in Fig.\ 1.  Note that the $d=3$ RFIM
one seems the furthest away from the MF Lorentzian of all the examples
for $T>T_c(H)$, whereas for $T>T_c(H)$ the pure $d=2$ case is
further than the $d=3$ RFIM, though both are very far
from MF behavior.

The critical parameters for the RFIM with $x>x_e$ are shown in
Table 4.  Certainly more effort is needed to complete the
experimental entries and to find reconciliation between the
simulation and experimental results.  Note that some
sets of exponents from the simulations violate\cite{nue97}
the Rushbrooke scaling relation in Eq.\ 3.

\section{The Vacancy Percolation Threshold Concentration}

The regions of low temperature nonequilibrium behavior and
equilibrium behavior have been shown to be separated at relatively
small $H$ by a nearly vertical sharp boundary at $x \approx x_e= 0.755$
in $\rm {Fe}_x\rm {Zn}_{1-x}\rm {F}_2$ using Monte
Carlo studies\cite{bb00jap}.
Figure 7 shows simulations on lattices  of size $2\times L^3 $ with $L=64$
modeled closely after $\rm {Fe}_x\rm {Zn}_{1-x}\rm {F}_2$.
Hysteresis
is observed upon FC and ZFC for $x<x_e$ but not above.
The hysteresis for $x<x_e$ increases for larger lattices or
slower thermal cycling, showing that it is not simply
an artifact of the simulations not being run long enough.
The concentration dividing equilibrium and nonequilibrium
behavior is very close to or equal to the
vacancy percolation threshold concentration $x_e=0.755$.  Apparently,
the percolation vacancy
structure facilitates the formation of domain walls.

To further investigate this boundary, recent experiments have been
done\cite{bybf00} on a sample with $x=0.76$, just above $x_e$.
No evidence of domains has been observed for small $H$.
Since earlier experiments\cite{kjbr85} for $x=0.72$ in
$\rm {Fe}_x\rm {Zn}_{1-x}\rm {F}_2$ gave clear evidence
for domain formation, including a reversal of the Bragg intensity
curvature just below $T_c(H)$ upon ZFC, the boundary must
be $0.72<x_e<0.76$, in agreement with the MC simulations.
There have been no theoretical studies reported explaining
the existence or nature of this boundary.

\section{The Current Situation and Outlook}

The recent measurement of the equilibrium critical behavior in 
the random-field Ising model has side-stepped the great difficulties
encountered in the interpretation of data below the transition
that are obfuscated by
nonequilibrium phenomena in many studies at lower concentrations.
Certainly, there is wide agreement that
a phase transition exists.  Although experiments for
$x>x_e$ are much more difficult since the random-field region
is very narrow, they are being done.  Interestingly, the
experimental results are not in agreement with much of the
theory and simulation results, unlike the REIM and pure Ising
model.  A reliable characterization of some aspects of the
$d=3$ RFIM universality class behavior remains to be completed.
The phase diagram of $\rm {Fe}_x\rm {Zn}_{1-x}\rm {F}_2$
has proven to be quite rich in detail.
An important area of the phase diagram for which
a good understanding is being developed is the large random
field limit for $x_p<x<x_e$.
With the the progress being made along these two lines of inquiry,
a rather complete experimental characterization of the RFIM
in dilute antiferromagnets seems near at hand.

Recent work has been supported by Department of Energy Grant
No. DE-FG03-87ER45324.

\begin{figure}
\psfig{figure=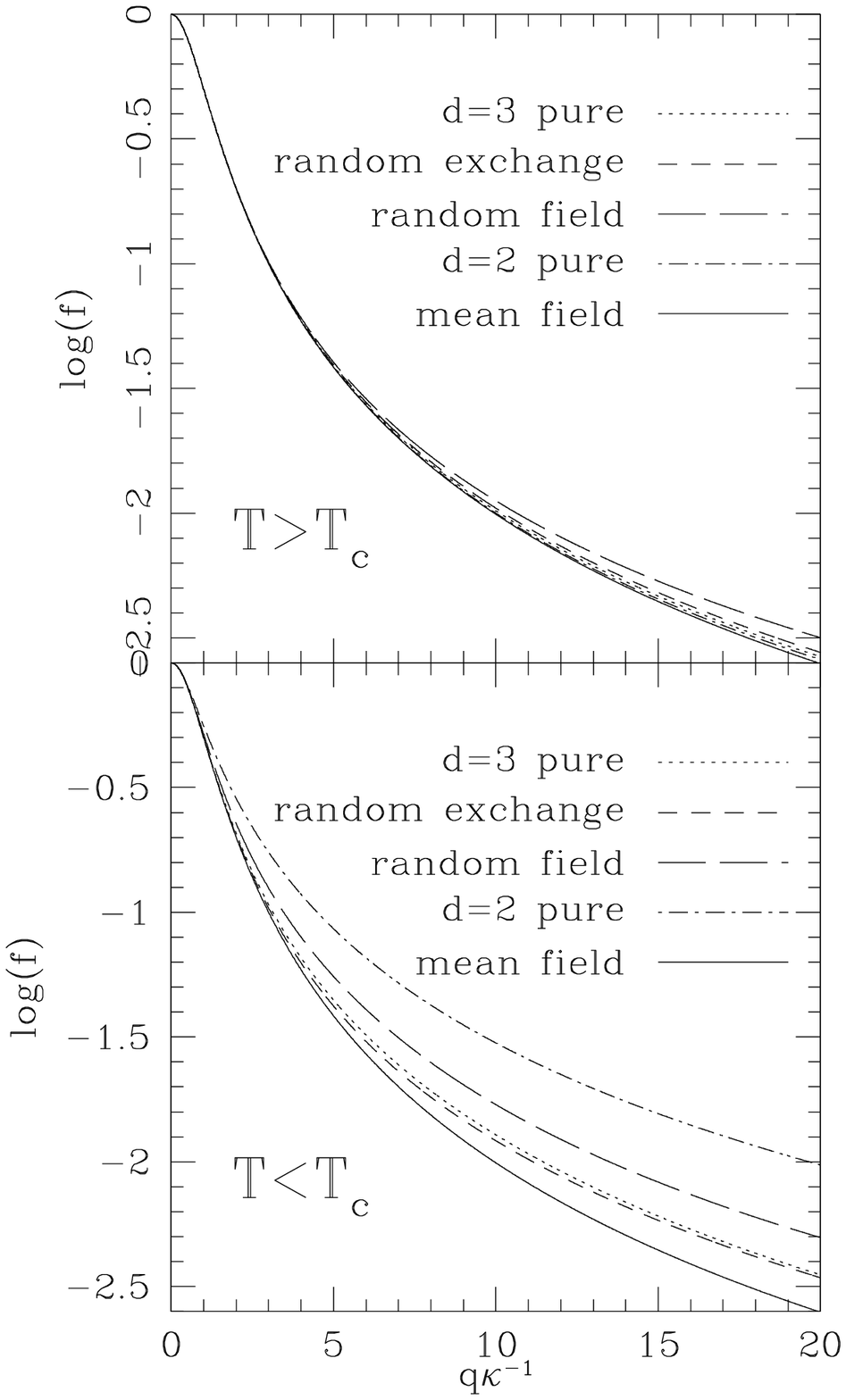,height=6.0in}
\caption{
A comparison of the logarithm of the scaling functions
$f(q/\kappa)$ versus $q/\kappa$ for different models (See Eq.\ 1).
The pure cases are from approximate expressions from numerical
studies\cite{tf75,fb67}.
The REIM and RFIM are determined from the experiments.
Note that the corrections to the MF
equation are largest below the transition and are very significant
for the pure $d=2$ and random-field $d=3$ cases.
}
\end{figure}

\newpage

\begin{figure}
\psfig{figure=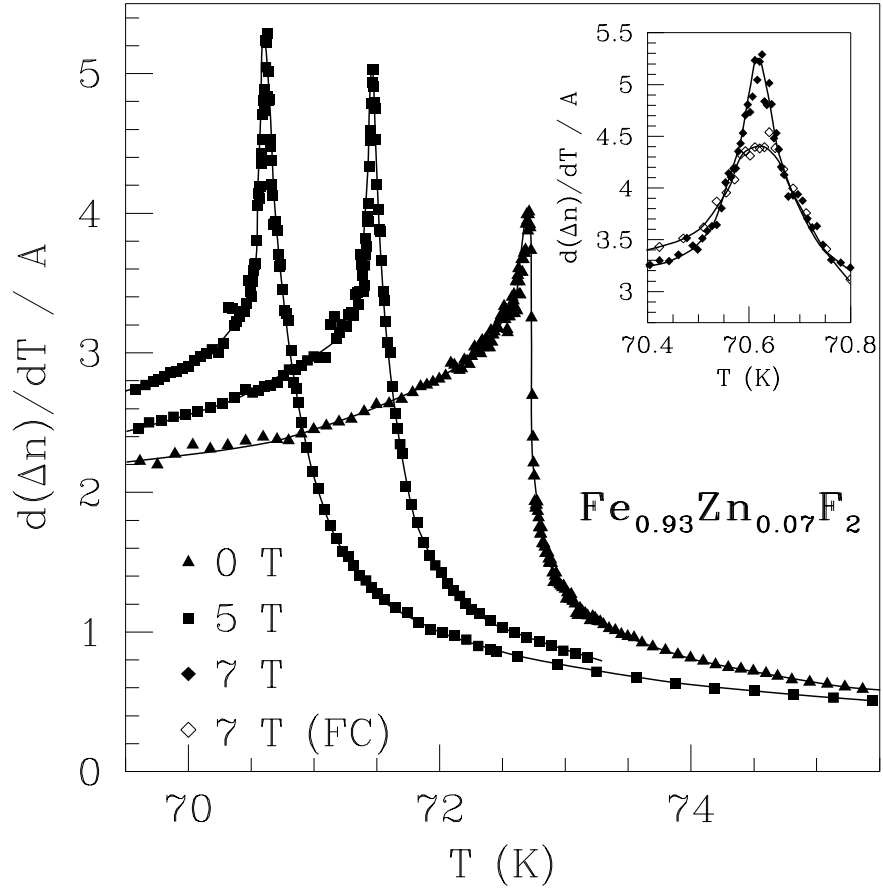,height=5.0in}
\caption{$C_m$ vs.\ $T$ for $\rm Fe_{0.93}Zn_{0.07}F_2$
at $H=7$~T as determined using the birefringence technique.
The inset shows the FC data.  There appears to be a tiny
hysteresis very close to the transition, perhaps a consequence
of random-field activated dynamics.
}
\end{figure}

\newpage

\begin{figure}
\psfig{figure=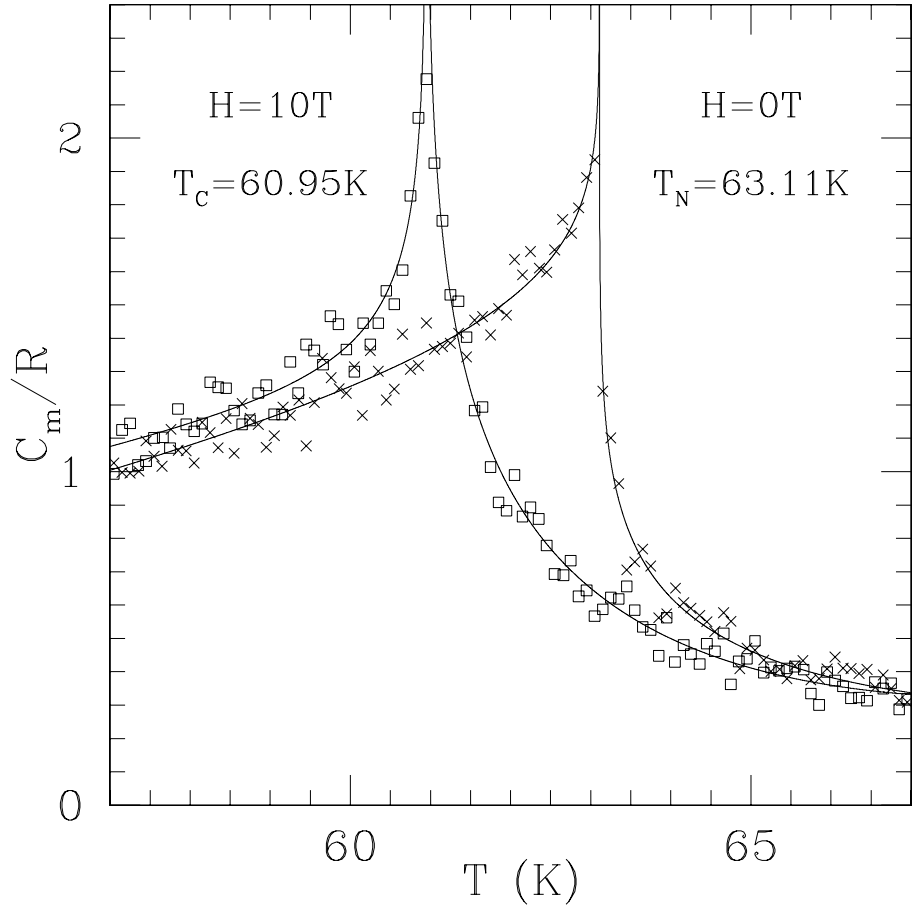,height=5.0in}
\caption{$C_m$ vs.\ $T$ from Monte Carlo simulations modeled after
the $\rm Fe_{0.8}Zn_{0.2}F_2$ system.
The simularity with the data is striking, though not all Monte Carlo
simulations yield a sharp peak in $C_m$.
}
\end{figure}

\newpage

\begin{figure}
\psfig{figure=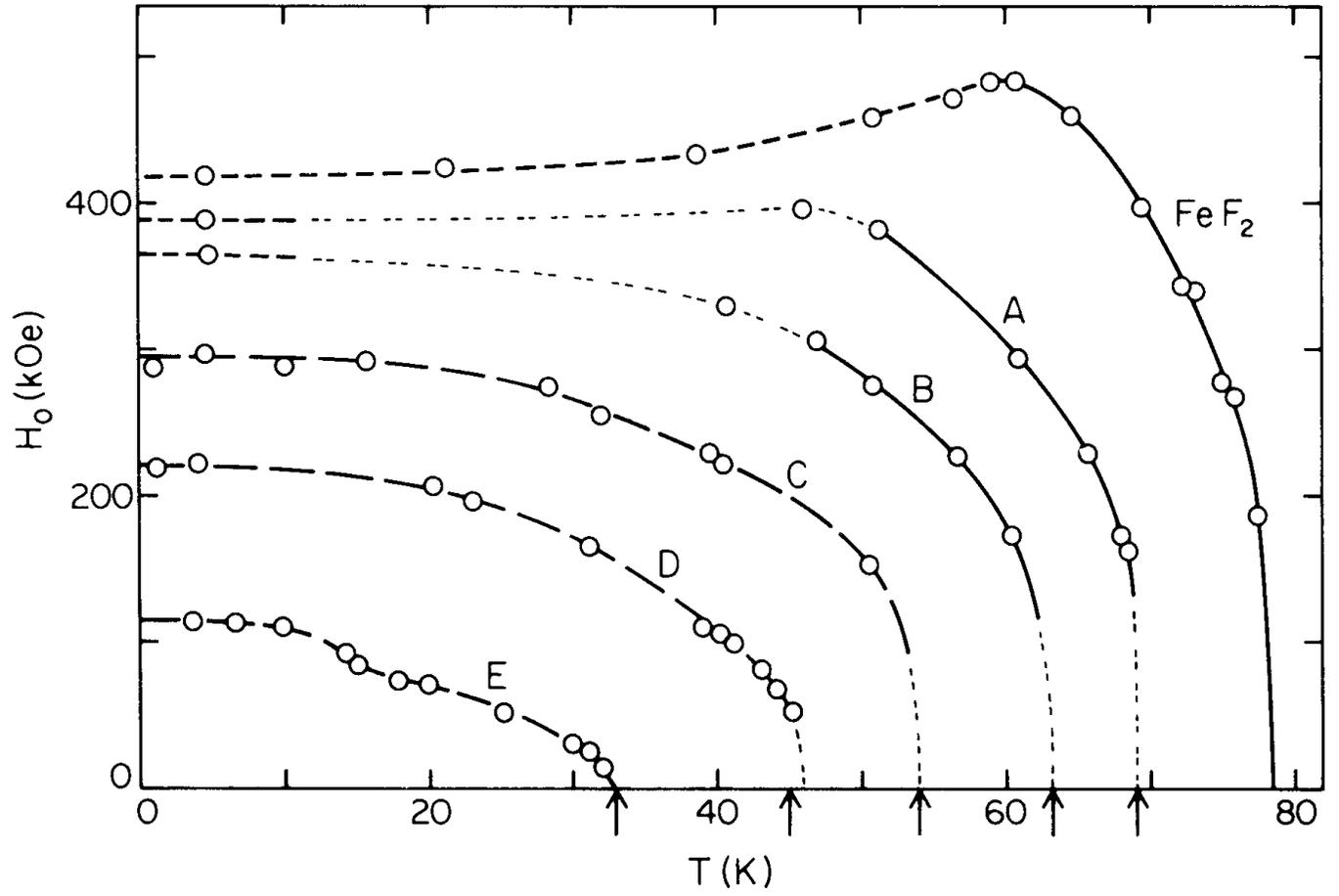,height=5.0in}
\caption{
The $H-T$ phase diagram for $Fe_xZn_{1-x}F_2$ measured in pulsed
magnetic fields.  The concentrations for the alphabetic labels
are the same as in Fig.\ 5.
}
\end{figure}

\newpage

\begin{figure}
\psfig{figure=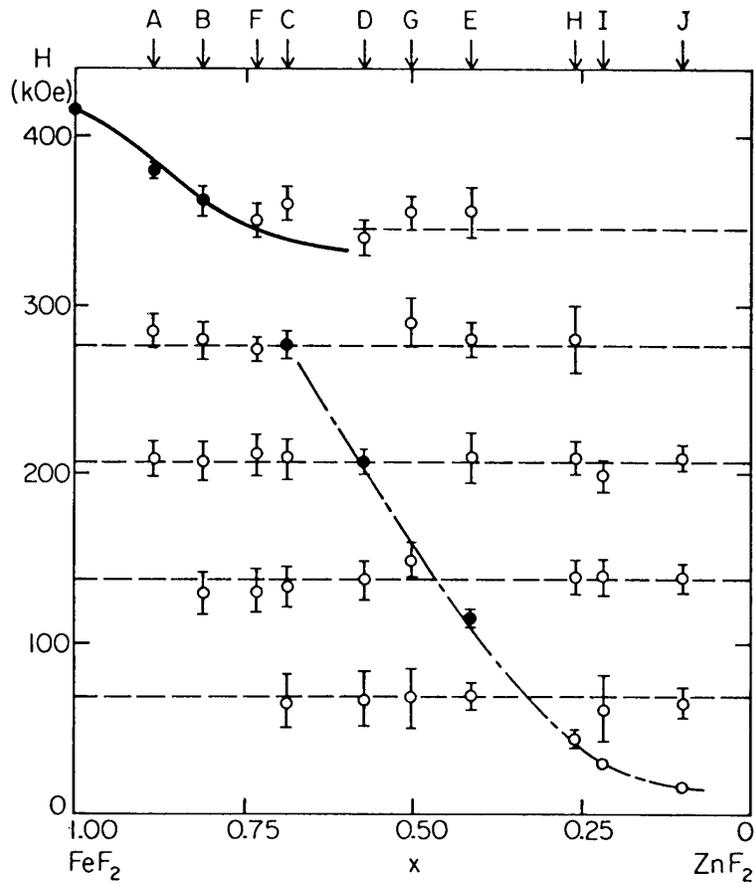,height=5.0in}
\caption{
Low temperature spin flips and phase boundary for
$Fe_xZn_{1-x}F_2$ as a function of $x$.  Note that the
phase boundary behavior is quite different for $x<x_e$ and $x>x_e$.
}
\end{figure}

\newpage

\begin{figure}
\psfig{figure=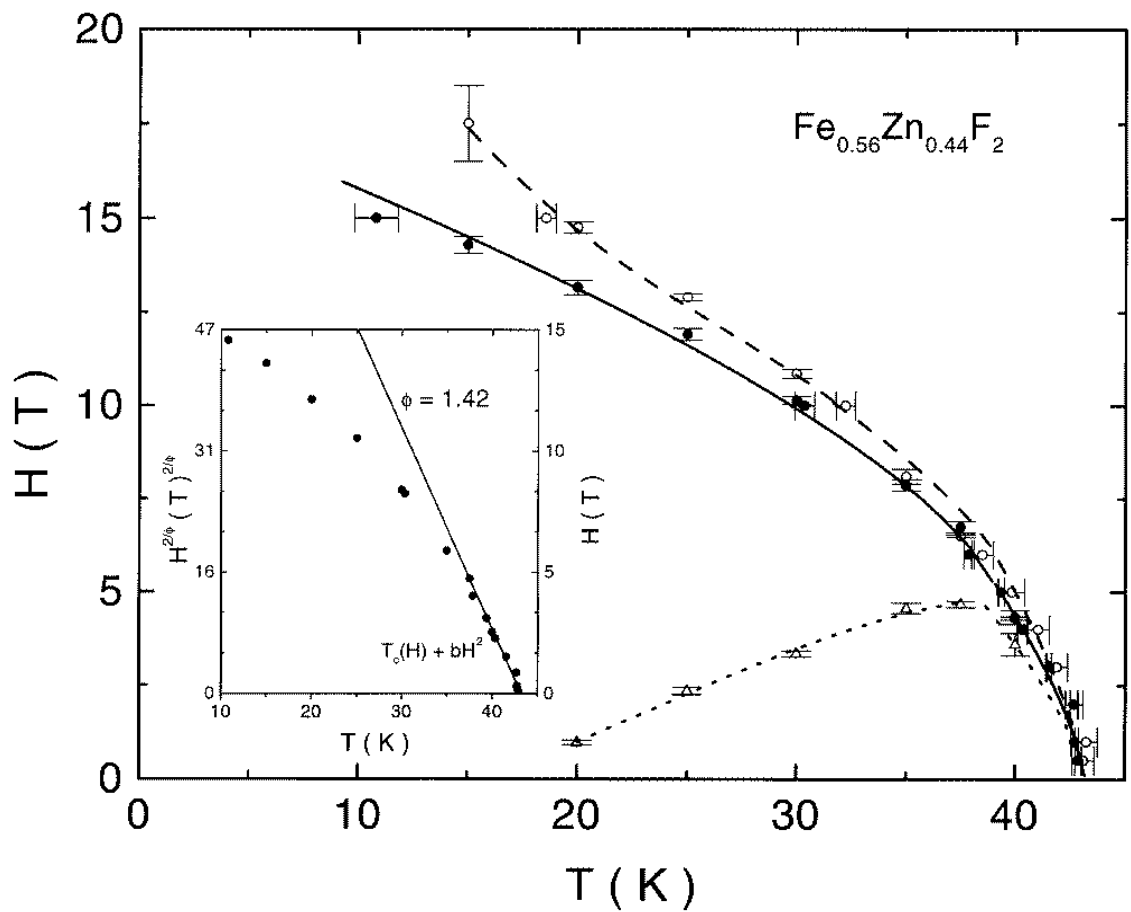,height=5.0in}
\caption{
The $H-T$ phase diagram for $x=0.56$ showing the upper equilibrium boundary,
the phase boundary and a lower equilibrium boundary.
}
\end{figure}

\newpage

\begin{figure}
\psfig{figure=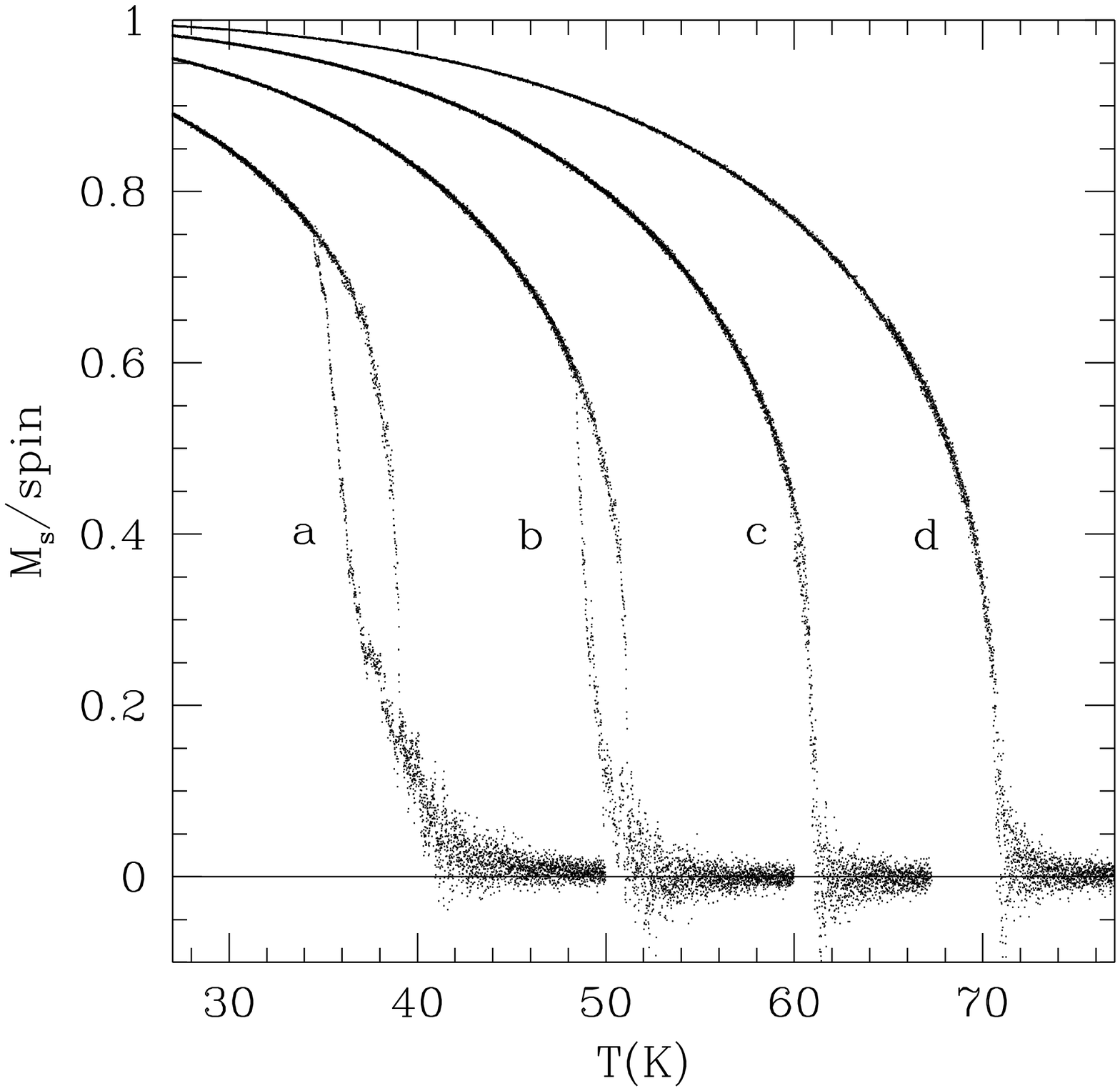,height=5.0in}
\caption{
Monte Carlo simulation data for the staggered magnetization versus $T$
for magnetic concentrations $0.5$, $0.6$, $0.7$ and $0.8$.  The ZFC and FC
procedures exhibit hysteresis for the lower concentrations, which only
gets worse for slower runs or larger lattices.  No hysteresis is observed
for the higher concentrations.
}
\end{figure}


\begin{thebibliography}{999}
\bibitem{rkjr88}The only measurement is using the dilation
technique at lower concentration where the system is not
in equilibrium by C. A. Ramos, A. R. King, V. Jaccarino and S. M. Rezende,
J. de Phys.\ {\bf 49}, C8-1241 (1988).
\end{thebibliography}
\end{document}